%% file: main.tex
\documentclass[sigconf]{acmart}
\usepackage{amsmath}
\AtBeginDocument{%
  \providecommand\BibTeX{{%
    \normalfont B\kern-0.5em{\scshape i\kern-0.25em b}\kern-0.8em\TeX}}}

\copyrightyear{2023}
\acmYear{2023}
\setcopyright{acmlicensed}\acmConference[CAMS 2023]{1st Workshop on Computer Architecture Modeling and Simulation}{October 28, 2023}{Toronto, Canada}
\acmBooktitle{1st Workshop on Computer Architecture Modeling and Simulation (CAMS 2023), October 28, 2023, Toronto, Canada}
\acmPrice{15.00}
\acmDOI{10.1145/3589236.3589244}
\acmISBN{979-8-4007-0776-6/23/02}

\begin{document}

\title{Analyzing and Improving Hardware Modeling of Accel-Sim}

\author{Rodrigo Huerta}
\email{rodrigo.huerta.ganan@upc.edu}
\orcid{0000-0003-0052-7710}
\affiliation{%
  \institution{Universitat Politècnica de Catalunya}
  \city{Barcelona}
  \country{Spain}
}

\author{Mojtaba Abaie Shoushtary}
\email{mojtaba.abaie@upc.edu}
\orcid{0000-0003-2377-6939}
\affiliation{%
  \institution{Universitat Politècnica de Catalunya}
  \city{Barcelona}
  \country{Spain}
}

\author{Antonio González}
\email{antonio@ac.upc.edu}
\orcid{0000-0002-0009-0996}
\affiliation{%
  \institution{Universitat Politècnica de Catalunya}
  \city{Barcelona}
  \country{Spain}
}

\renewcommand{\shortauthors}{Huerta and Abaie, et al.}

\begin{abstract}
\par
GPU architectures have become popular for executing general-purpose programs. Their many-core architecture supports a large number of threads that run concurrently to hide the latency among dependent instructions. In modern GPU architectures, each SM/core is typically composed of several sub-cores, where each sub-core has its own independent pipeline.

\par
Simulators are a key tool for investigating novel concepts in computer architecture. They must be performance-accurate and have a proper model related to the target hardware to explore the different bottlenecks properly.

\par
This paper presents a wide analysis of different parts of Accel-sim, a popular GPGPU simulator, and some improvements of its model. First, we focus on the front-end and developed a more realistic model. Then, we analyze the way the result bus works and develop a more realistic one. Next, we describe the current memory pipeline model and propose a model for a more cost-effective design. Finally, we discuss other areas of improvement of the simulator.
\end{abstract}



\maketitle

\input{1.Introduction}
\input{2.Proposal}
\input{3.Evaluation_Methodology}

\input{4.Results}
\input{5.Future_Work}

\input{6.Conclusion}

\begin{acks}
This work has been supported by the CoCoUnit ERC Advanced Grant of the EU’s Horizon 2020 program (grant No 833057), the Spanish State Research Agency (MCIN/AEI) under grant PID2020-113172RB-I00, and the ICREA Academia program. 
\end{acks}

\bibliographystyle{ACM-Reference-Format}
\bibliography{sample-base}

\end{document}

%% file: 1.Introduction.tex
\section{Introduction}

\par
GPU architectures have become popular for executing general-purpose programs \cite{Burtscher2012} in addition to graphics workloads. These architectures have many cores, also known as Streaming Multiprocessor (SM) or Compute Units in Nvidia and AMD terminology respectively, that share an L2 cache. GPUs' programming model is based on having a vast amount of threads that are arranged into Cooperative Thread Arrays (CTA). Each CTA is mapped onto an SM. Threads in a CTA can easily synchronize and share data through a configurable scratchpad memory inside each SM, typically called Shared Memory. Once a kernel (a task executed in a GPU) is launched, CTAs are assigned to SMs. Threads in a CTA are grouped into sets (typically of 32 or 64 threads each) referred to as warps (also known as wavefronts). All threads in a warp execute in parallel in a lockstep mode, known as SIMT (single instruction multiple threads) execution mode. In modern architectures, each core is normally subdivided into different sub-cores (usually 4) \cite{voltaPaper}, \cite{TuringPaper}, \cite{AmperePaper}, \cite{AdaPaper}, \cite{HopperPaper} and the warps of each CTA are distributed among them. Each sub-core has its own independent L0 instruction cache and pipeline.

\par
\begin{figure}[ht]
  \centering
  \includegraphics[trim={0.6cm 0.6cm 0.6cm 0.6cm},clip,width=8cm]{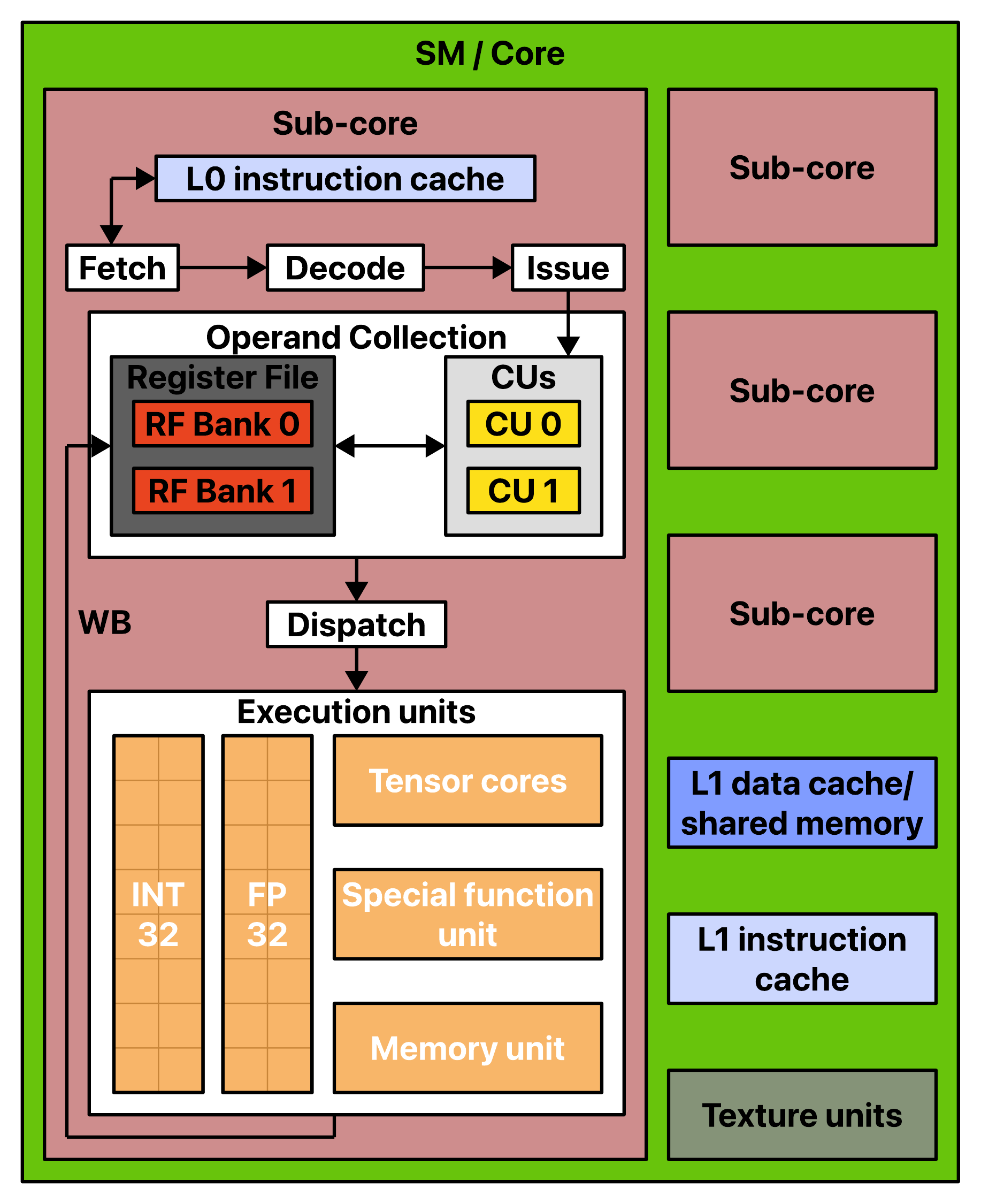}
  \caption{SM architecture.}
  \label{fig:TuringSM}
\end{figure}

In \autoref{fig:TuringSM}, we can see an image illustrating the most important parts of an SM of modern GPU architectures.

\par
The first stage of a typical GPU pipeline is the Fetch, where a round-robin scheduler selects a warp with empty Instruction Buffer slots to start a fetch request of a few (e.g., two) consecutive instructions from the L0 instruction cache. When the request is completed, the instructions are decoded in the Decode stage and placed into the Instruction Buffer of the corresponding warp.

\par
In the issue stage, a warp among all the eligible ones is selected to issue its oldest instruction. An example of a widespread issue policy in the literature is Greedy Then Oldest (GTO) \cite{GTO}. A warp is eligible to be scheduled if it has at least one instruction in the Instruction Buffer, and the oldest instruction does not depend on previously executed instructions pending to be finished. In this stage, instructions check its dependencies before being issued. A well-known approach for handling dependencies is using Scoreboards. However, other alternatives, such as a hardware-compiler co-designed approach, are used by some modern GPU architectures.

\par
Once an instruction is issued, it is placed in a Collector Unit (CU), and waits until all its source register operands are retrieved. Each sub-core register file has multiple banks with a few (e.g., two) ports per bank, allowing for multiple accesses in a single cycle at low cost. An arbiter deals with the possible conflicts among several petitions to the same bank. When all source operands of an instruction are in the CU, the instruction goes to the dispatch stage, where it is dispatched to the proper execution unit (e.g., memory, single-precision, special function) whose latencies differ depending on the type of unit and instruction. Once the instruction reaches the write-back stage, it writes its result in the register file.

\par
In order to explore new ideas, simulators have been widely used in computer architecture, and GPGPUs are no exception. NVIDIA has exposed part of the process of creating their warehouse simulator, NVIDIA Architectural Simulator (NVArchSim or NVAS) \cite{NVAS}. As for the academic side, we have different alternatives. One of them is MGPUSim \cite{mgpusim}, which is a multi-gpu simulator modeling the AMD GCN 3 micro-architecture and supporting virtual memory. An alternative simulator is the Accel-Sim \cite{accelsim} framework, a cycle-accurate state-of-the-art trace-driven simulator supporting CUDA applications trying to resemble NVIDIA modern architectures.

\par
This work focuses on improving the modeling of three main components inside the SM and sub-cores in the Accel-Sim framework: the front-end,  the result and the memory pipeline.

\par
The rest of this paper is organized as follows. In \autoref{sec:proposal} we present the different enhancements done to the simulator to have a more realistic model. In \autoref{sec:evaluationmethodology}, we describe the evaluation methodology that is later used in \autoref{sec:results} to analyze the effects of the proposed modifications. Then, we continue discussing in \autoref{sec:futureWork} other improvements that we plan to investigate in the future. Finally, we conclude in \autoref{sec:conclusions}.

%% file: 2.Proposal.tex
\section{Improvements to Accel-Sim}\label{sec:proposal}

\par
This section discusses the SM modeling problems we found in the simulator and how we solved them. First, we start talking about the front-end in \autoref{subsec:problemsfrontend}. Then, we focus on the result bus modeling in \autoref{subsec:problemsresultbus}. Finally, we improve the memory execution pipeline model in the \autoref{subsec:problemsldstunit}.

\subsection{Front-end}\label{subsec:problemsfrontend}

\par
The latest architectures of NVIDIA have a design of SMs with four sub-cores or processing blocks in NVIDIA terminology \cite{voltaPaper}, \cite{TuringPaper}, \cite{AmperePaper}, \cite{AdaPaper}, \cite{HopperPaper}. Each sub-core has a decode and a fetch unit that accesses each sub-core's private L0 instruction cache. In the upper memory level, there is a L1 instruction cache shared by the four sub-cores. As is reported by Zhe Jia et al. \cite{dissectingVolta} \cite{dissectingTuring} and Barnes et al. \cite{subcoreBalancing}, warps are distributed between sub-cores with the following formula: $\textrm{sub-core\_id} = \textrm{warp\_id} \% 4$.

\par
\begin{figure}[ht]
  \centering
  \includegraphics[trim={0.6cm 0.6cm 0.6cm 0.6cm},clip,width=6cm]{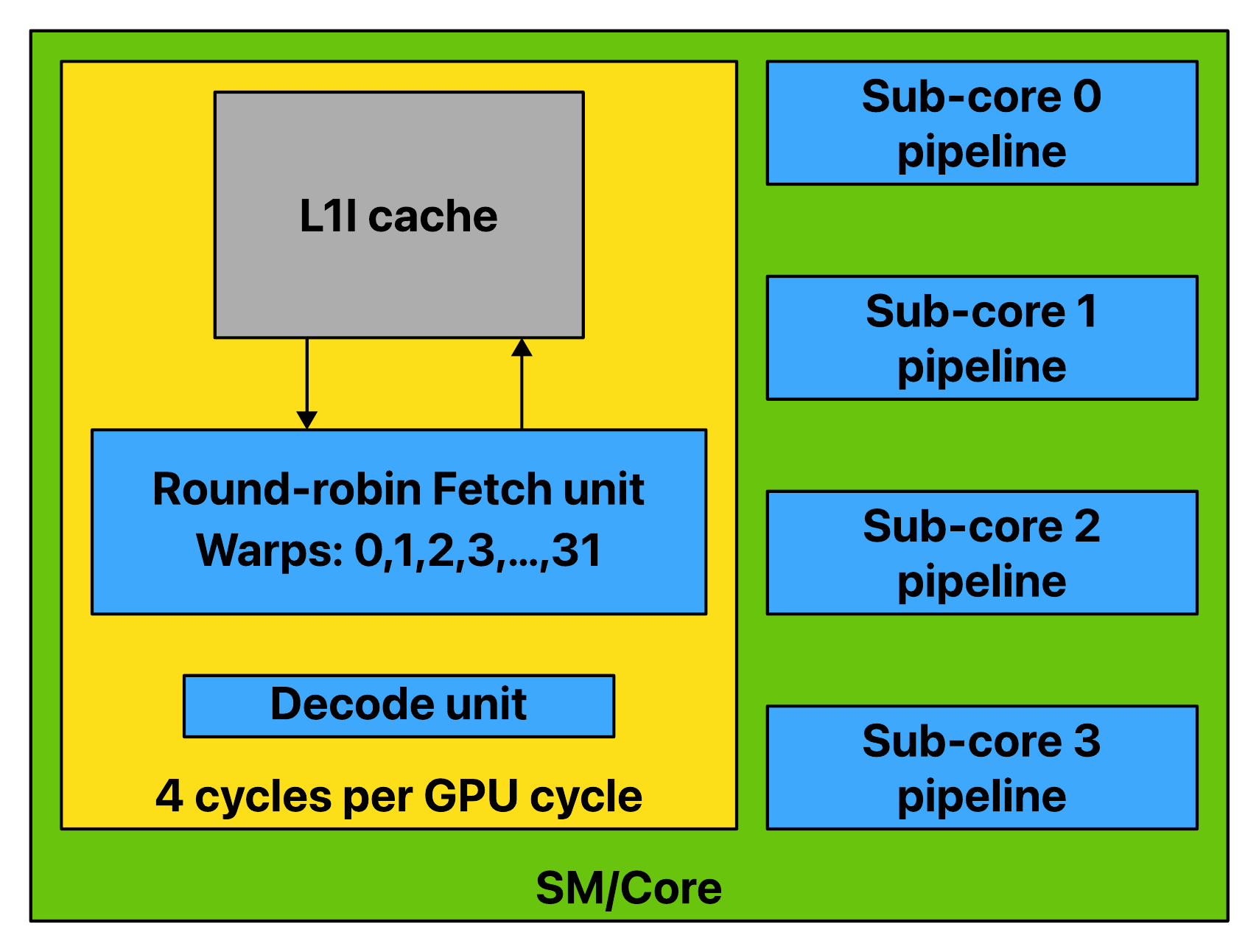}
  \caption{Current Accel-Sim front-end.}
  \label{fig:oldFrontend}
\end{figure}

In \autoref{fig:oldFrontend}, we can see a block diagram of what is modeled in Accel-Sim. As it can be seen, there are no private L0 instruction caches per subcore, and the split among sub-cores starts after the decode stage, meaning there is only one fetch and decode unit for the whole SM instead of four. In spite of that, the pipeline can be kept fully occupied because the fetch and decode stages are done four times per cycle, but this model does not reflect modern architectures, and it requires a much more costly solution, since the shared instruction cache must be able to serve four different access per cycle. Furthermore, there are cases where the fetching and decoding of the instruction is happening in just one cycle, which is incorrect.

\par
\begin{figure}[ht]
  \centering
  \includegraphics[trim={0.6cm 0.6cm 0.6cm 0.6cm},clip,width=8cm]{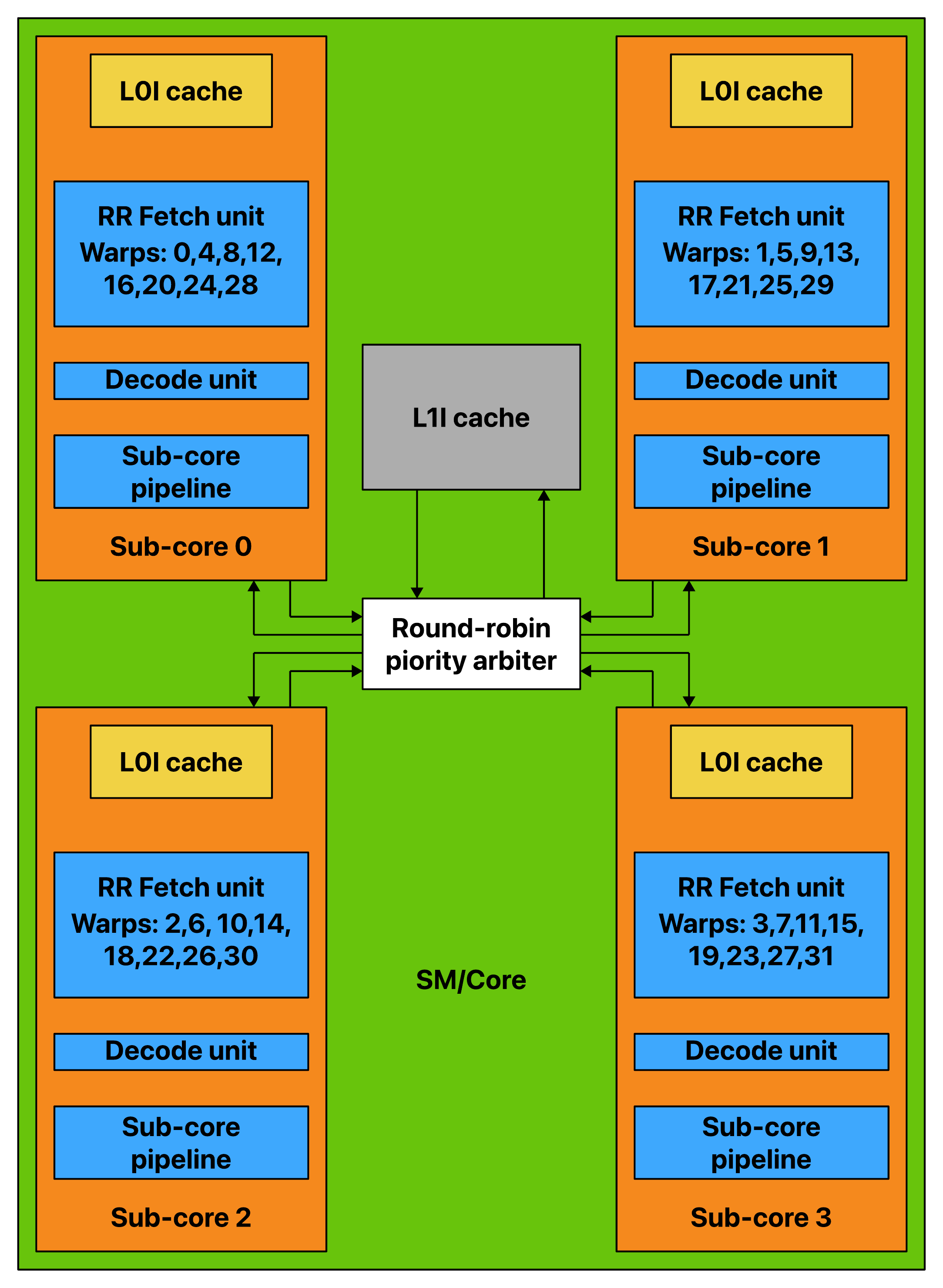}
  \caption{Proposed front-end.}
  \label{fig:newFrontend}
\end{figure}

We have modified the front-end model of Accel-Sim to model the architecture depicted in \autoref{fig:newFrontend}, which is based on the sub-core architecture of modern GPUs. As it can be seen, there is an  L0 instruction cache, a fetch unit and a decode unit private to each sub-core. Besides, there is a round-robin priority arbiter for handling the requests from different sub-cores to the L1 instruction cache.

\par
Another improvement over current Accel-Sim is regarding the way instructions are packed and stored in the intruction cache. The current simulator fetches two instructions per warp four times per cycle. The simulator is trace-driven and it packs together instructions that may belong to different cache lines and instructions that do not have consecutive addresses because branches (e.g., instruction at PC \texttt{0x100} is packed with instruction at PC \texttt{0x460}). Furthermore, instructions belonging to different kernel codes have the same address, which makes the instruction cache to wrongly compute many accesses as hits when they should be misses. For instance, the first two instructions of each kernel are assumed to have the same memory addresses whereas in reality they are different. In our model, we have fixed this issues, by mapping instructions to cache lines in a correct manner, and giving different addresses to instructions of different kernels.

\par
In short, we have developed a more accurate/realistic model of the front-end by adding private L0 instructions caches, fetch and decode units per sub-cores, and have implemented a correct and realistic mapping of instructions to cache lines.

\subsection{Result bus}\label{subsec:problemsresultbus}

\par
During the dispatch of instructions with fixed execution latencies to the different execution pipelines, the availability of result buses to write-back the register file is checked. The result bus modeling in the original simulator is just searching for a free result bus with the instruction latency after being dispatched to the execution unit. However, it does not consider conflicts of register file banks at write-backs of fixed latency operations. So, we have added support to detect these conflicts and only allow the same number of instructions that go to the same register file bank finishing in a given cycle as the number of register bank ports is in the architecture. 

\par
It is remarkable that during this modeling of the result bus, we assume two register file ports as the latest architectures of NVIDIA are designed with two register file ports per register file bank as discussed by Zhe Jia et al. \cite{dissectingVolta} \cite{dissectingTuring} instead of only one. Moreover, each port can be used for a write-back or a read. So, up to two instructions with the same destination register file bank being executed in different execution units can be scheduled to finish in the same cycle. 

\subsection{Memory execution pipeline}\label{subsec:problemsldstunit}

\par
Modern GPUs based on sub-core partitioning have a memory pipeline in each sub-core, which accesses shared memory structures for the whole SM (L1 data cache, shared memory, texture cache, constant memory) \cite{voltaPaper}, \cite{TuringPaper}, \cite{AmperePaper}, \cite{AdaPaper}, \cite{HopperPaper}. Even though the behavior is unrevealed, it has the sense that each of these memory pipelines is in charge of calculating the memory addresses and coalescing before sending the requests to the memory structures.

\par
\begin{figure}[ht]
  \centering
  \includegraphics[trim={0.6cm 0.6cm 0.6cm 0.6cm},clip,width=8.5cm]{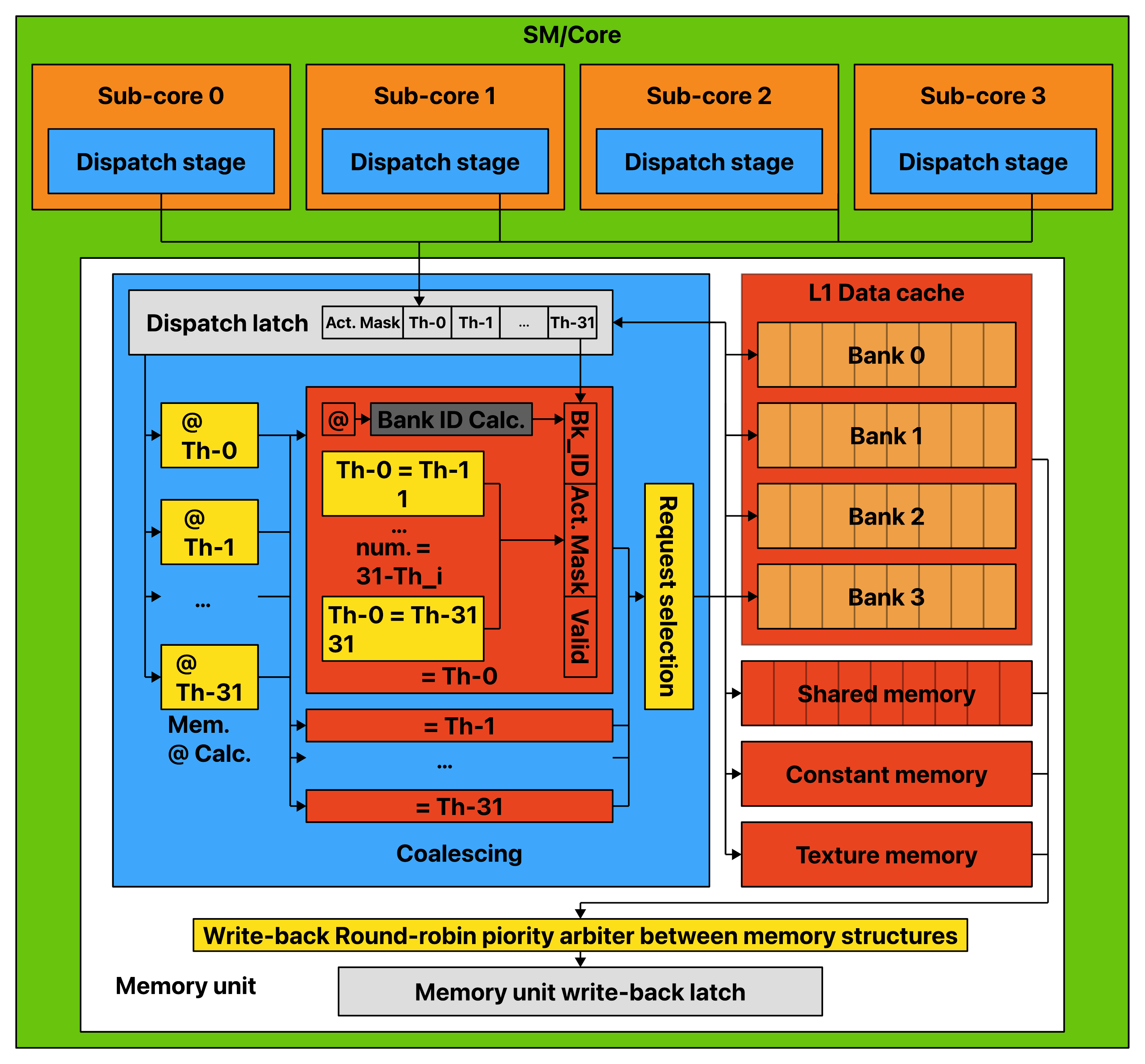}
  \caption{Memory execution pipeline of Accel-Sim.}
  \label{fig:oldMemoryPipeline}
\end{figure}

The Accel-Sim choice is to have a single memory pipeline for the whole SM instead of a dedicated unit per sub-core. This creates a problem because there is a single dispatch latch for the whole SM instead of one per sub-core. Instructions are maintained in this latch until all the requests have been sent to the desired memory structure. Satisfying all the requests of an instruction can lead to many cycle stalls depending on the degree of achieved coalescing and bank conflicts of the instructions requests. It prevents other memory instructions from progressing and stops other instructions from being issued (even from different sub-cores). As memory instructions are not dispatched, they are held in operand collector units (a limited resource inside the sub-core), so any instruction is prevented from being issued because there is no space in the operand collection stage. Moreover, address calculation, coalescing, and request selection are designed to be done in the same cycle, greatly increasing hardware requirements. As it is constrained to do all these tasks in a single cycle, the more hardware-hungry part is the coalescing. The reason is that each thread needs to know which threads have the same address, so the total number of address comparators is $C_{32,2} = 496$, which is huge. Furthermore, there is a single write-back latch for the whole SM, which may create some contention in case there are different memory structures with ready accesses simultaneously for different sub-cores. This became worse if the instruction in the write-back latch could not progress because register bank ports were already occupied. In \autoref{fig:oldMemoryPipeline}, we can see a graphical representation of the baseline simulator model for the memory pipeline.

\par
\begin{figure}[ht]
  \centering
  \includegraphics[trim={0.06cm 0.6cm 0.6cm 0.6cm},clip,width=8.5cm]{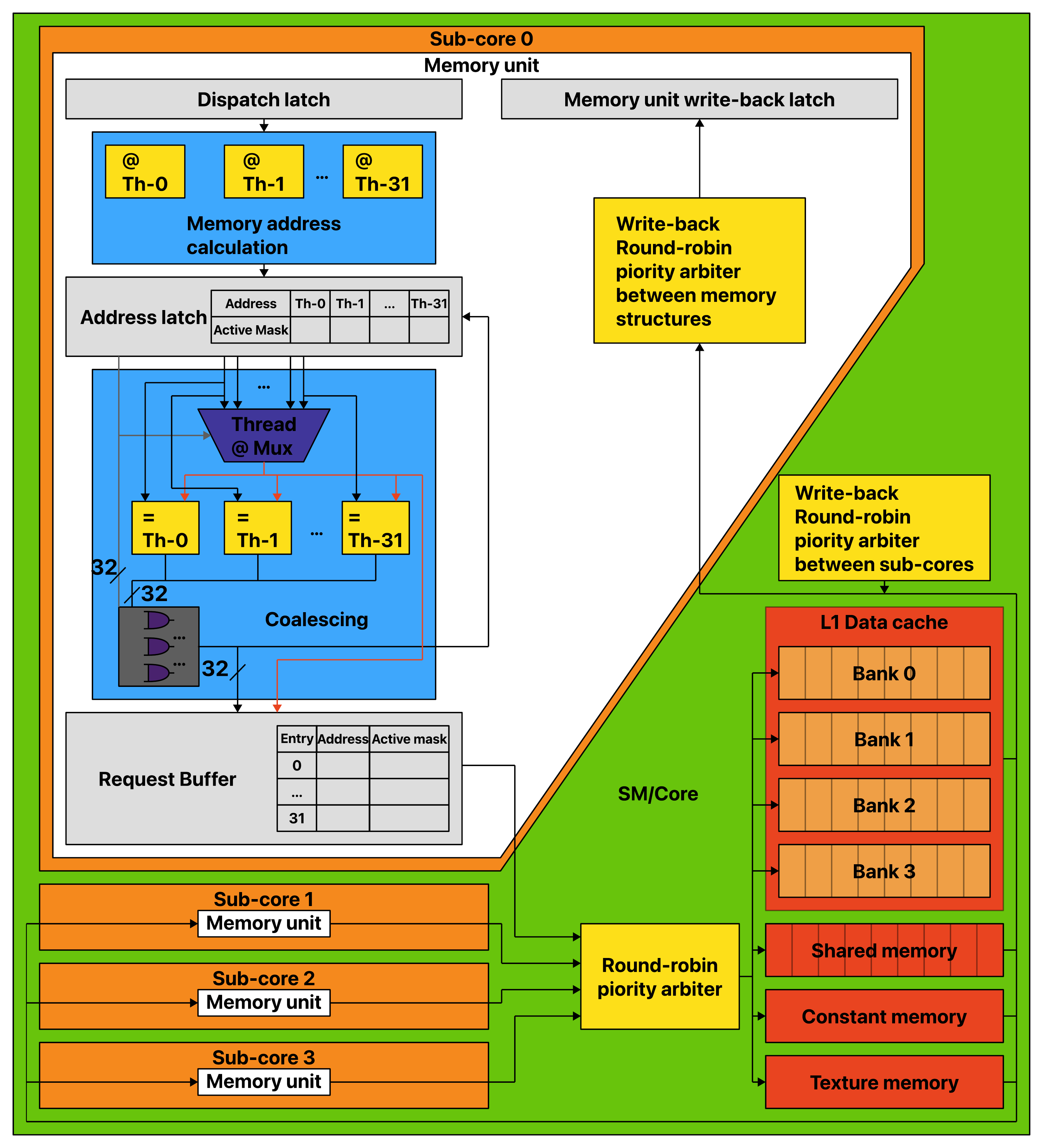}
  \caption{Proposed memory execution pipeline.}
  \label{fig:newMemoryPipeline}
\end{figure}

Regarding our model, we have incorporated a Memory pipeline unit in each sub-core that accesses the memory structures shared for the whole SM. The details of this unit of commercial GPUs are unknown; therefore, we have addressed an aggressive performance design that is reasonable, which might differ from commercial designs. We have modeled a pipeline inside this unit that splits the memory address calculation and coalescing in different cycles. An instruction is maintained in the address latch until all the memory addresses have been processed for being coalesced, but the instruction just remains in the dispatch latch one cycle unless the address latch is not empty. The number of cycles needed for coalescing will depend on the requests required for each warp. For example, if all the accesses go to the same cache block, it will need just a single cycle. Concerning the coalescing hardware, as it dedicates one cycle for each thread, the number of address comparators is just $32$. Once an address has been processed, it is stored in the request buffer. All the memory requests stored in that buffer are sent to the round-robin priority arbiter between sub-cores. This means that if there is a memory structure or a bank from L1D unused, it will be used for that request, even if that is not the first one generated by the coalescing unit. This differs from the current model from Accel-Sim, which only allows requests to progress in order of generation. Finally, we have a write-back latch for each sub-core with an arbiter to prioritize between different memory structures. Moreover, we include a write-back arbiter to give priority between sub-cores for popping accesses from the different memory structures. A depiction of this model can be found in \autoref{fig:newMemoryPipeline}.

%% file: 3.Evaluation_Methodology.tex
\section{Evaluation Methodology} \label{sec:evaluationmethodology}

\par
To evaluate how these improvements in the SM have affected the simulation, we have measured the speed-up and the Absolute Variation in Cycles (AVC) over the baseline simulator. We have used 42 benchmarks belonging to Rodinia 3.1 \cite{rodinia3}, Deepbench\cite{deepbenchWeb}, Parboil \cite{parboil}, Pannotia \cite{pannotia}, and ISPASS-2009 \cite{ispass} suites simulated under completion. 

\par
We have configured the Accel-Sim \cite{accelsim} simulation infrastructure with the trace execution mode to reassemble an NVIDIA RTX 2070 Super. The main configuration parameters in \autoref{tab:gpu_specs}. Moreover, we have extended the simulator to include the different fixes explained in \autoref{sec:proposal}.

\begin{table}
  \caption{GPU specification}
  \centering
  \label{tab:gpu_specs}
  \begin{tabular}{c c }
    \toprule
    Parameter & Value\\
    \midrule
    Clock & $1605$ $MHz$ \\
    SP/INT/SFU/Tensor Units per sub-core & 1/1/1/1\\
    Warps per SM & 32 \\
    Warp Width & 32 \\
    Number of registers per SM & 65536 \\
    Issue Scheduler policy & GTO \\
    Number of SMs & 40 \\
    Sub-cores per SM & 4 \\
    Number of Collector Units per sub-core & 2 \\
    L1 instruction cache size & $32$ KB \\
    L1 data cache size & $32$ KB \\
    Shared memory size & $64$ KB\\
    L2 cache size & $4$ MB \\
    Memory Partitions & 16 \\
    \toprule
    \multicolumn{2}{c}{Fixes} \\
    \midrule
    L0 instruction cache size & $16$ KB \\
    Max. Num. requests and replies of L0I & 1 \\
    Register file ports per bank & 2 \\
  \bottomrule
\end{tabular}
\end{table}

%% file: 4.Results.tex
\section{Results} \label{sec:results}

\par
In this section, we analyze the impact of the different incorporated fixes.

\par
Even these changes report a tiny change in performance, just a $0.25\%$ speed-up and a $3.67\%$ of AVC on average, there are significant changes if we look in detail at some benchmarks. The reason why the average speed-up is smaller than the AVC is because there are benchmarks that are gaining performance, and others are losing it. At the same time, AVC measures the absolute difference between the original version and the proposed one.

\par
\begin{figure}[ht]
  \centering
  \includegraphics[width=8.8cm]{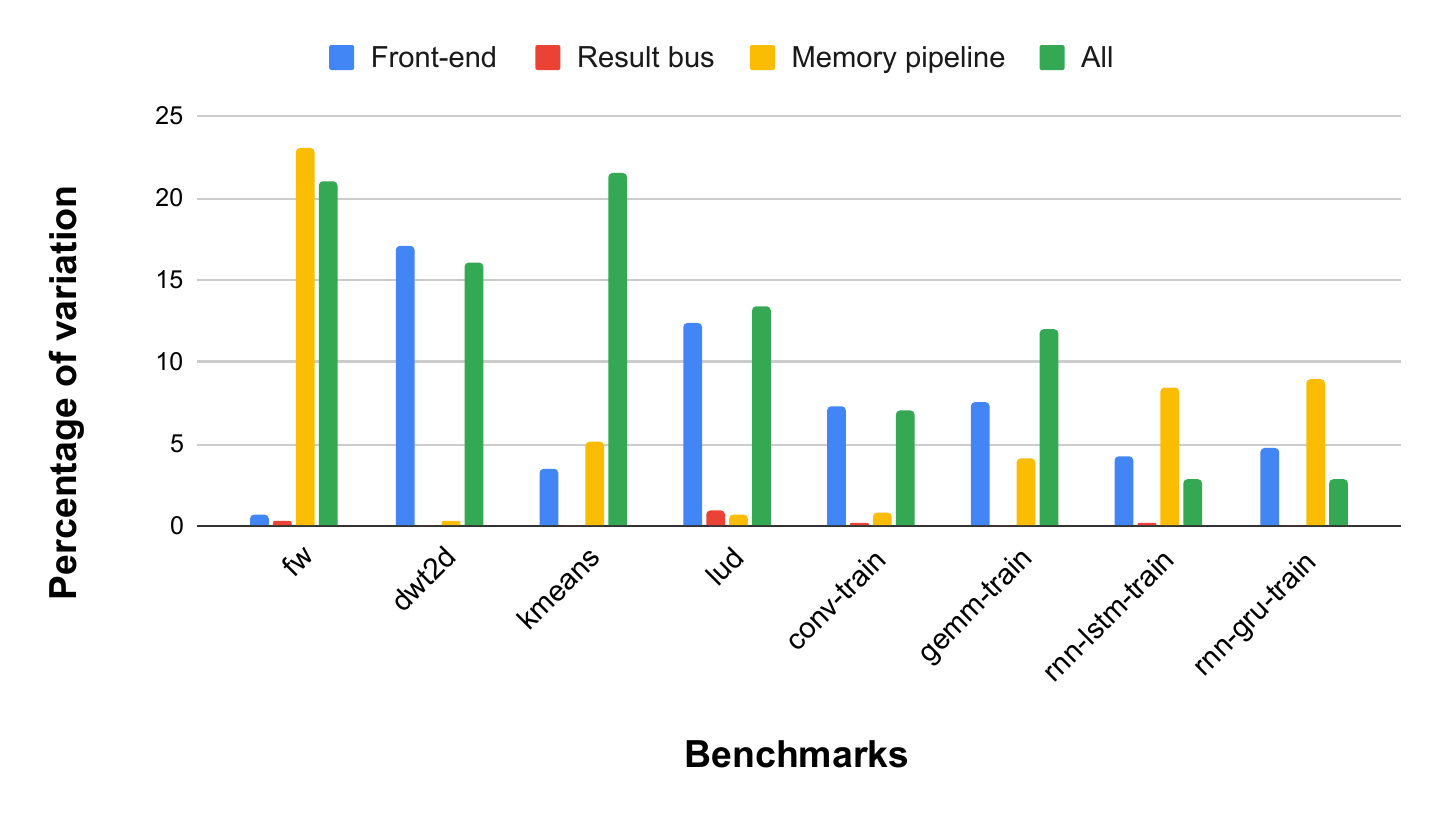}
  \caption{Absolute Variation of Cycles of the most significant applications.}
  \label{fig:variation}
\end{figure}

In \autoref{fig:variation}, we can see the impact of the different improvements in some significant benchmarks. The all configuration may have less impact than separate fixes because they create differences in cycles in opposite directions. Nevertheless, some benchmarks such as \textit{gemm-train} are affected in the same direction, so including all the changes produces an effect of $12\%$.

\par
In this figure, we can see that the result bus modeling is the change that is affecting the less. 

\par
However, the changes in the memory pipeline are more noticeable in more benchmarks, where the most significant case is \textit{fw} reaching a $23\%$ of AVC regarding the memory pipeline and $21\%$ including all the changes. 

\par
\begin{figure}[ht]
  \centering
  \includegraphics[width=8.8cm]{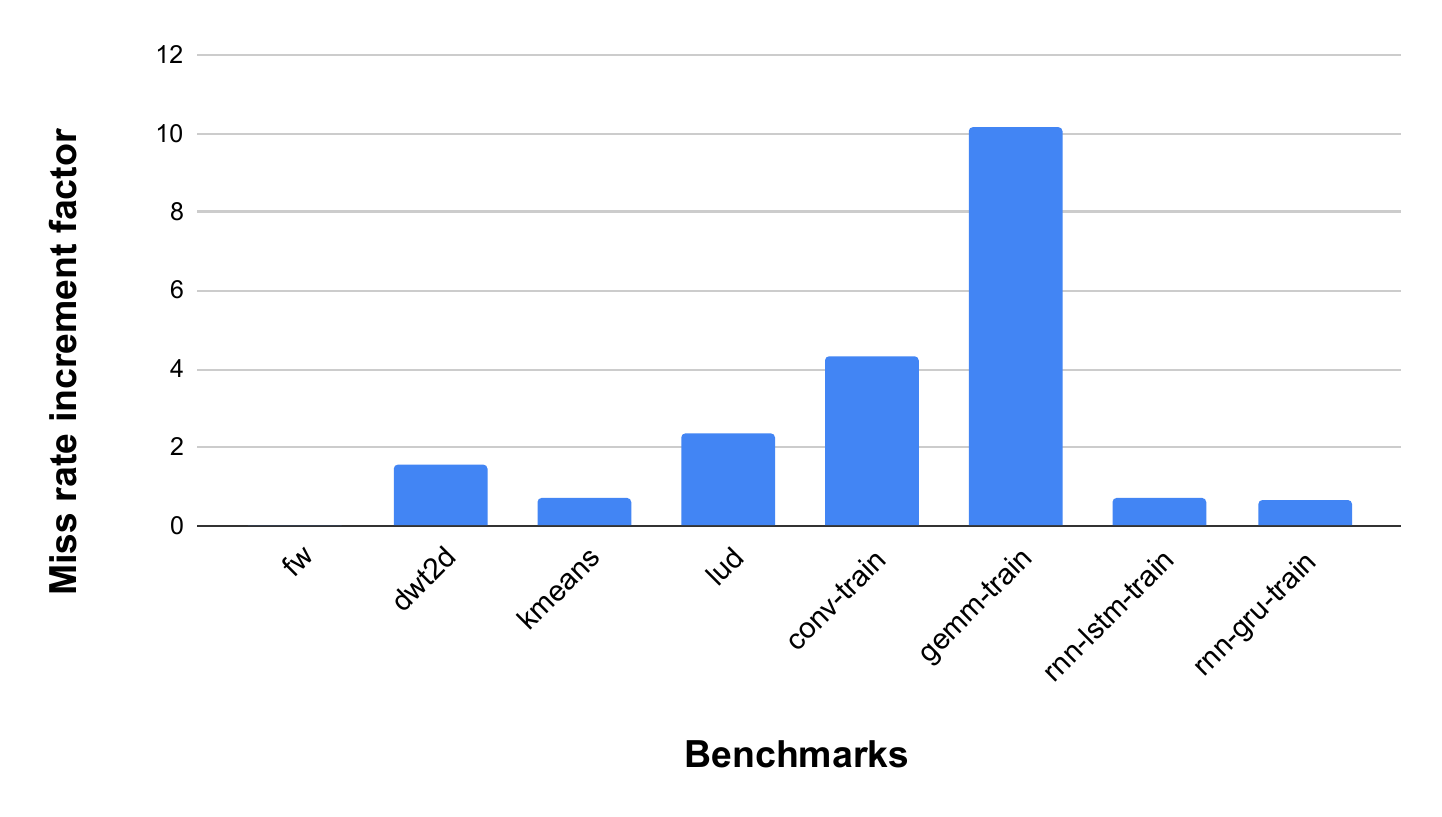}
  \caption{Miss ratio.}
  \label{fig:missRatioVariation}
\end{figure}

About the modifications affecting the front-end, it is affecting a lot \textit{dwt2d} ($17\%$) and \textit{lud} ($12.36\%)$. The main reason behind these variations in performance is the different enhancements regarding instruction caches. To illustrate it, we can see in \autoref{fig:missRatioVariation} the miss rate increment factor of the first level instruction cache. 

\par
If we compare \textit{gemm-train} against \textit{dwt2d} and \textit{lud}, it requires more than $27.4X$ and $7.8$ cycles respectively to complete the execution. This is why even though the miss rate increment factor is more significant in \textit{gemm-train}, it has less effect than in shorter benchmarks.

%% file: 5.Future_Work.tex
\section{Future work} \label{sec:futureWork}

\par
Apart from the improvements in modeling that we have presented in this paper, others will be necessary to improve the simulator and open the door to new research areas in GPUs. We will start discussing what can be done inside the GPU cores, and then we will move outside the SM.

\subsection{Inside SM} \label{subsec:insideSM}

\par
First of all, as reported by Mishkin et al. \cite{warHazards}, the simulator is not handling WAR dependencies correctly. Even though the issue of instructions is in-order, the dispatch from collector units to the execution pipelines is out-of-order. It could be the case that two instructions of the same warp are in collector units simultaneously, and the younger one with WAR hazard is dispatched earlier than the older one. Besides, this problem is not frequent, as reported by Mishkin et al. \cite{warHazards}, hardware must support these cases not to incur errors. We believe that commercial GPUs are not suffering from WAR hazards, but as discussed by Zhe Jia et al. \cite{dissectingVolta} \cite{dissectingTuring}, the management of dependencies of modern GPUs seems that is detecting instruction dependencies without scoreboards. However, we lack detailed documentation on how the hardware supports it.

\par
Regarding the Operand Collection stage and the register file of GPUs, it is known that the latest architectures of NVIDIA are designed with two register file ports per register file bank as discussed by Zhe Jia et al. \cite{dissectingVolta} \cite{dissectingTuring} instead of only one. In Accel-Sim, they have modeled this feature focusing on the throughput of reading operands. However, the way that is handled is by repeating twice the Operand Collection stage per cycle. This means that the allocation of collector units, dispatch of ready collector units, and arbitration of reads are repeated twice per cycle. Moreover, it is assumed that each collector unit has unlimited ports for reading the operands, which would mean having a huge crossbar to support the worst case. It was one in previous architectures, but this parameter is unknown in the current ones.

\par
Additionally, the register file caching system for instruction operands is not present. Instead, the number of register banks is increased, which may be effective for having an equivalent contention to commercial hardware. Due to this lack of modeling, it is difficult to know if novel proposals such as Malekeh \cite{Malekeh} are beating industry designs.

\par
Furthermore, the tracking usage of registers across the execution is uncompleted. The tracer tool only captures the usage of regular registers, but registers such as predication, uniform, and uniform predicate are not detected. Furthermore, some instructions, such as tensor ones, use two registers for some operands even though only one of them is captured or indicated in the binary, so investigating this conduct will grant the community more opportunities to investigate problems in operand collection. 

\par
Finally, as the state-of-the-art mode of the simulator is trace-driven, it focuses on capturing executed instructions. In addition, control flow instructions like \texttt{BMOV} use special registers that can have dependencies with general purpose registers. Therefore, the current model is not enough to correctly analyze many applications' control flow behavior. This complicates analyzing and creating new micro-architectural and compiler proposals regarding this topic.

\subsection{Ouside SM} \label{subsec:outsideSM}

\par
Outside of the SM, other topics can be improved in this simulation tool. First of all, the NOC between SMs and memory partitions is plain in the simulator without taking into account the hierarchy of TPCs and GPCs, which seems to be the approach followed by the industry, as can be seen in this patent \cite{NOCpatent}. This hierarchy modeling will allow us to analyze new opportunities, including the new Hopper architecture features such as thread-block cluster and distributed shared memory between SM inside a GPC \cite{HopperPaper}. Finally, adding virtual memory and multi-GPU support will be great for analyzing new kinds of trendy workloads.

%% file: 6.Conclusion.tex
\section{Conclusion} \label{sec:conclusions}

\par
In this paper, we propose different improvements to the Accel-Sim framework with the purpose of having a more real simulator. We have focused on the front-end modeling (supporting sub-core split and better modeling of instruction caches). Also, we have a better design of the result bus. Then, we explored how the baseline simulator represents the memory pipeline and how we think it should be modeled to have a more feasible approach. Finally, we have compared all these modifications against the baseline simulator to show how they affect the performance of benchmarks.